\documentclass[12pt]{article}
\usepackage{amsmath}
\usepackage{graphicx}

\title{Interaction between dust grains near a conducting wall}
\author{A.M. Ignatov\thanks{e-mail: aign@fpl.gpi.ru} \\
\textit{\small Institute of General Physics, } \\ \textit{\small
 Russian Academy of Sciences, ul. Vavilova 38, Moscow, 119991
Russia}  }

\date{}
\begin{document}
\maketitle

\begin{abstract}
The effect of the conducting electrode on the interaction of dust
grains in a an ion flow is discussed. It is shown that two grains
levitating above the electrode at the same height may attract one
another. This results in the instability of a dust layer in a
plasma sheath.
\end{abstract}

\section{Introduction}

A good deal of experiments in dusty plasma physics are performed
with aerosol grains suspended in the plasma sheath.
 A negatively charged dust grain in the sheath levitates above a
 horizontal electrode
 due to the
 balance of two forces: the gravity force  directed downwards and the
 sheath electric field that pushes the grain upwards.
The near-sonic or the supersonic  ion flow in the sheath creates
the wake field downstream of the grain.  Since the latter is
confined by a certain Mach cone, it is commonly accepted that the
wake field affects the motion of grains, which are  situated
downstream only. The usual assumption is that the intergrain
potential is smooth in the horizontal direction, \textit{i.e.,}
two grains levitating at the same height repel one another via the
screened Coulomb potential. The structure of the wake field and
the grain interaction in an ion flow were studied in details
[1-5]. The asymmetric interaction of vertically aligned grains was
also observed experimentally \cite{melzer}.

Analytical theory and computer simulations cited above assumed
that the plasma density is constant and the influence of the
conducting electrode is negligible. Although both assumptions
evidently fail under  conditions of  a real plasma sheath, taking
into account the plasma non-uniformity seems to be a very
difficult problem. In order to estimate the influence of the
electrode upon grain interaction, here I use the zeroth
approximation, which however, seems  the only one treatable
analytically.

 In the present paper, the following simple model is accepted.
  Let there be the monoenergetic ion stream entering a conducting
electrode (or a grid) located at the horizontal plane, $z=0$. The
stream velocity, $u$, exceeds the ion thermal velocity but it is
much less than the electron thermal velocity.  Electrons are
Boltzmann distributed. Two problems are addressed: first, how does
the electrode modify the interaction between two grains levitating
at the same height and, second, how this affects the spectrum of
dust acoustic waves propagating along a single dust layer.

 \section{Intergrain interaction}

 The electrostatic potential
produced by a point charge, $Q$, located at
$\mathbf{r}=\mathbf{r}_0$ is given by the solution of the Poisson
equation

\begin{equation}\label{poisson} \Delta \hat{\varepsilon} \varphi(\mathbf{r})=-4\pi Q
\delta(\mathbf{r}-\mathbf{r}_0), \end{equation}
where $\hat{\varepsilon}$ is the operator of the static dielectric
permittivity of an ambient plasma.  Within the accepted model, the
spatial Fourier transform of $\hat{\varepsilon}$ is given by

\begin{equation}\label{eps} \varepsilon(\mathbf{k})=1-\frac{\omega_i^2}{(k_z u +i
0)^2}+\frac{k_{D}^2}{k^2},\end{equation}
where $u$ is the velocity of the ion flow, $\omega_i$ is the ion
plasma frequency and $k_D$ is the inverse electron Debye length.
The ion flow is parallel to the $z$ axis and directed downwards,
$u<0$.

In unbounded plasma, the natural boundary condition for
Eq.~(\ref{poisson}) is $\varphi|_{r\to \infty}\to0$. It is
convenient to express the solution to Eq.~(\ref{poisson}) in terms
of its Green function, \textit{i.e.,} $\varphi(\mathbf{r})=Q
G^0(\mathbf{r}-\mathbf{r}_0)$, where the Fourier transform of
$G^0(\mathbf{r})$ with respect to the transverse coordinate
$\mathbf{r}_\perp=(x,y)$ is

\begin{equation}\label{gi0} G^0_{k_\perp}(z)=\int\frac{d k_z}{2\pi} e^{i k_z z}
\frac{4\pi}{k^2 \varepsilon(\mathbf{k})}.\end{equation}
Here $\mathbf{k}_\perp=(k_x,k_y)$ stands for  the transverse
components of a wave vector. With the dielectric function given by
Eq.~(\ref{eps}), the zeros of denominator in Eq.~(\ref{gi0}) are

\begin{eqnarray} k_z&=&\pm k_D q-i0,\nonumber\\ k_z&=&\pm i k_D \kappa,
\label{zeros}\end{eqnarray} where

\begin{eqnarray}\label{q} q^2&=&\frac12 \left\{
\sqrt{(q_\perp^2+1-\mu^2)^2+4\mu^2 q_\perp^2}-q_\perp^2 +\mu^2 -1
\right\},\\
\kappa^2&=&\frac12 \left\{ \sqrt{(q_\perp^2+1-\mu^2)^2+4\mu^2
q_\perp^2}+q_\perp^2 -\mu^2 +1 \right\},
\label{kappa}\end{eqnarray}
 $q_\perp=k_\perp/k_D$ is the normalized transverse wave vector,
and $\mu=\omega_i/k_Du$ is the inverse Mach number.

The integral in Eq.~(\ref{gi0}) is readily evaluated resulting in
\begin{equation}\label{g0}
G^0_{k_\perp}(z)=\frac{2\pi}{k_D}\frac{1}{q^2+\kappa^2}\left[\kappa
e^{-\kappa k_D |z|} + 2 \theta(-z) q \sin\left( q k_D z\right)
\right].\end{equation}

The spatial structure of the potential is recovered by the Fourier
transform of the Green function (\ref{g0}) with respect to
$k_\perp$.  The first term in parenthesis in expression (\ref{g0})
gives rise to the Debye-H\"uckel potential distorted by the ion
flow, while the second term represents the wake field situated
downstream of the charge [1-3]. The opening of the Mach cone
confining the wake and the field structure inside it depend on the
stream velocity.

Now we turn to the evaluation of the electric potential of a
charge located near a conducting wall. Let the wall be situated at
$z=0$ plane, while the charge is placed above it, $z_0>0$. Then
the potential is given by the solution of Eq.(\ref{poisson})
supplemented with the boundary condition $\varphi|_{z=0}=0$. As it
is well-known from electrostatics, we can make allowance for this
boundary condition by introducing the surface charge density,
$\sigma_{k_\perp}^{ind}$, induced at the conducting surface. Then,
the potential of a unit charge is written as

\begin{equation}\label{gp1} G_{k_\perp}(z,z_0)=G_{k_\perp}^0(z-z_0)+
G_{k_\perp}^0(z)\sigma_{k_\perp}^{ind}.\end{equation}

Taking into account the boundary condition,
$G_{k_\perp}(0,z_0)=0$, we find the surface charge density
$\sigma_{k_\perp}^{ind}$ and, finally,

\begin{equation}\label{gp} G_{k_\perp}(z,z_0)=G_{k_\perp}^0(z-z_0)-
\frac{G_{k_\perp}^0(z)G_{k_\perp}^0(-z_0)}{G_{k_\perp}^0(0)}.\end{equation}

One may doubt whether description a bounded dispersive medium  in
terms of the response function of an unbounded medium is
justifiable. However, more accurate and lengthy calculations give
the same result. The physical reason is that there are no ions
reflected by the conducting wall within the present model. The
mathematical reason is that Eq.~(\ref{poisson}) actually masks a
set of partial differential equations with two real
characteristics directed downwards.

Of particular interest for the following is the interaction
potential of two charges placed at the same height, $z_0$. Since

\begin{equation}\label{ge} G_{k_\perp}(z_0,z_0)=\frac{4\pi}{k_D} \frac{e^{-\kappa
a}}{\kappa^2+q^2}\left[\kappa \sinh(\kappa a)+q \sin(qa)\right],
\end{equation} where $a=k_D z_0$,  the normalized potential is

\begin{equation}\label{potn} w(\rho)=2 \int\limits_0^\infty q_\perp dq_\perp\;
J_0(q_\perp \rho)\frac{e^{-\kappa a}}{\kappa^2+q^2}\left[\kappa
\sinh(\kappa a)+q \sin(qa)\right],\end{equation}
where $G(r_\perp, z_0,z_0)=k_d w(\rho)$ and $\rho=k_d r$. The
asymptotic behaviour of $w(\rho)$ is determined by Eq.~(\ref{ge})
at $q_\perp\to 0$. The latter depends essentially on whether  the
the ion flow is supersonic or subsonic. In the case of the
supersonic flow, $\mu<1$, the roots (\ref{q},\ref{kappa}) at
$q_\perp\to0$ are approximated as

\begin{equation}\label{supsr} q\approx q_\perp \frac{\mu}{\sqrt{1-\mu^2}},\quad
\kappa\approx \sqrt{1-\mu^2},\end{equation} and the leading term
of the asymptotic expansion of $w(\rho)$ (\ref{potn}) is

\begin{equation}\label{wsup} w(\rho)|_{\rho\to\infty} \sim \frac{2}{\mu}
(1-\mu^2)^{3/2} e^{-a\sqrt{1-\mu^2}} \frac{\theta(a \mu -\rho
\sqrt{1-\mu^2})}{\sqrt{a^2
\mu^2-\rho^2(1-\mu^2)}}+O\left(e^{-r}\right).\end{equation}

This expression may be interpreted as a mirror reflection of the
wake field produced by the grain at $0<z<z_0$. More detailed
numerical investigation of the potential (\ref{wsup}) shows that
$w(\rho)$ is always positive for $\mu<1$.

Quite another behaviour is observed for the case of the subsonic
flow, $\mu>1$. The roots (\ref{q},\ref{kappa}) at $q_\perp\to0$
are now

\begin{equation}\label{subsr} q\approx \sqrt{\mu^2-1}  ,\quad \kappa\approx  q_\perp
\frac{\mu}{\sqrt{\mu^2-1}},\end{equation}

while  the potential  at infinity behaves like

\begin{equation}\label{wsub} w(\rho)|_{\rho\to\infty}\sim  \frac{2 a \mu}{\mu^2-1}
\sin\left(a\sqrt{\mu^2-1}\right) \frac1{\rho^3}.\end{equation}

The most important distinction between the expressions
(\ref{wsup}) and (\ref{wsub}) is that in the subsonic regime the
potential is attractive if

\begin{equation}\label{attr} \sin\left(a\sqrt{\mu^2-1}\right)<0.\end{equation}

\begin{figure}
\centerline{
\includegraphics{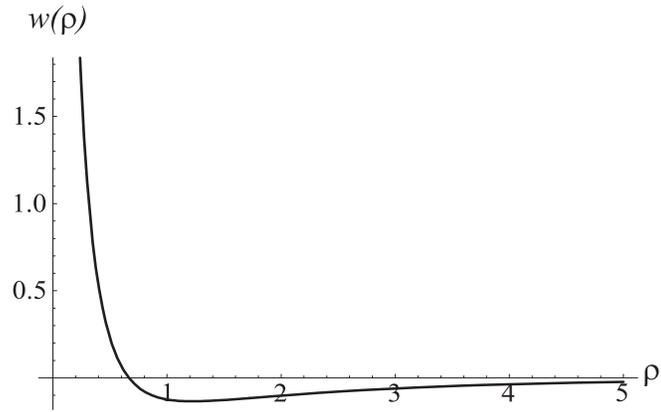}}
\caption{\label{fig1} The distribution of the potential in the
transverse direction.   $\mu=2$, $a=2.7$. }
 \end{figure}

\begin{figure}\centerline{
\includegraphics{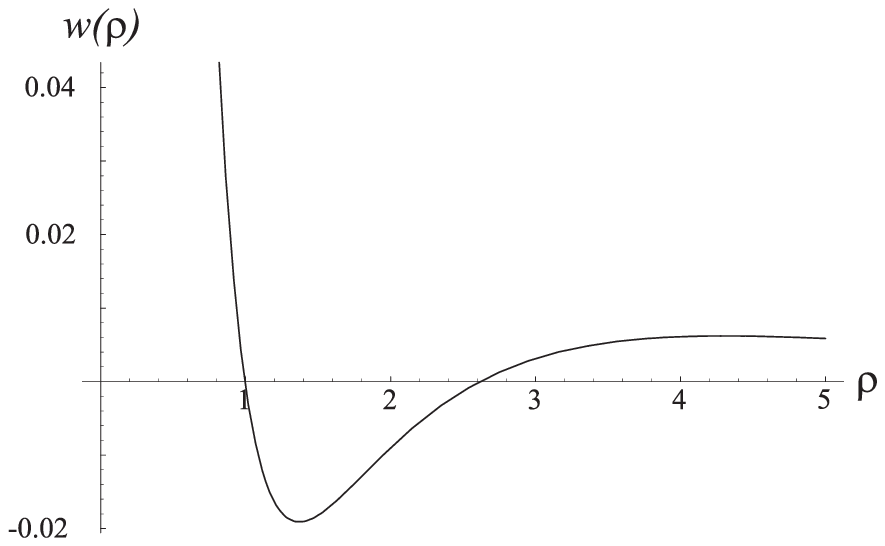}}
\caption{\label{fig2} The distribution of the potential in the
transverse direction.   $\mu=2$, $a=4$. }
 \end{figure}

The numerically evaluated  example of the potential (\ref{wsub})
demonstrating the attraction is depicted in Fig.~\ref{fig1}. It
should be pointed out that the inequality (\ref{attr}) guarantees
the long-scale attraction between grains.  With the opposite
inequality imposed on $a$ and $\mu$, the potential is repulsive at
large distance but the attractive branch may appear at smaller
scales, as shown in Fig.~(\ref{fig2}).

Evidently, the existence of the attractive branch of the
interparticle interaction may result in formation of various
patterns and clusters even in the absence of the external
potential well confining  grains in the horizontal direction.
Also, the even distribution of grains in a dust layer may become
unstable. The latter possibility is discussed in the next section.

\section{Dust layer}

Now consider a two-dimensional gas consisting  of dust grains
hovering over the conducting electrode. Ignoring intergrain
correlations, the linearized  equations of motion are written as

\begin{eqnarray}\label{cont}
\frac{\partial \sigma}{\partial t}&+&\nabla_\perp (\sigma_0
\mathbf{v})=0,\\
 \label{euler}\frac{\partial \mathbf{v}}{\partial t}&=&
 -\nu \mathbf{v}-\frac{Q^2}M k_d \nabla_\perp\int
 d\mathbf{r}_\perp^\prime w(k_D(\mathbf{r}_\perp-\mathbf{r}_\perp^\prime)),\end{eqnarray}
where $\sigma_0$ is the unperturbed value of the surface density
and
 $\sigma$ is the density perturbation. Here I consider horizontal
 motions only, \textit{i.e.,} $v_z=0$. The term $-\nu\mathbf{v}$ in
 Eq.~(\ref{euler}) corresponds to the grain friction on an ambient
 neutral gas. The intergrain interaction is described by the
 potential   given by Eqs.~(\ref{potn}). Although the grain
 charge, $Q$, generally depends on the ambient plasma parameters,
 for simplicity we ignore its variability.

 Assuming that all quantities are proportional to $ \exp(-i \omega t +i
 \mathbf{k}_\perp \mathbf{r}_\perp)$ we easily get the dispersion relation
 for the gas oscillations:

 \begin{equation}\label{disp}
 \omega (\omega+i\nu)=g_d \frac{k_\perp^2}{2\pi} G_{k_\perp}(z_0,z_0),\end{equation}
 where $g_d=2\pi Q^2 \sigma_0/M$.

\begin{figure}\centerline{
\includegraphics{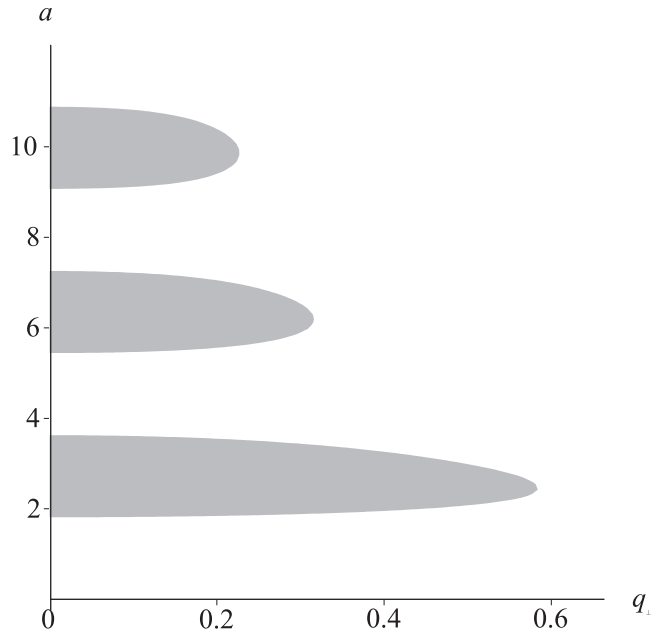}}
\caption{\label{fig3}Instability regions in the $(q_\perp, a)$
plane. $\mu=2$.  }
\end{figure}

This expression describes dust sound waves in the continuous
medium approximation. In the long-wave limit, $q_\perp\to0$, the
dispersion relation is of the form

\begin{equation}\label{displ}
 \omega (\omega+i\nu)=\frac{g_d k_d q_\perp^2}{\sqrt{|1-\mu^2|}}
   \begin{cases}
 e^{-a\sqrt{1-\mu^2}}\sinh( a\sqrt{1-\mu^2}), &
 \mu<1\\
\sin( a\sqrt{\mu^2-1}), &\mu>1
 \end{cases}
 \end{equation}
Evidently, the layer is unstable, \textit{i.e.,} $\textrm{Im\,}
\omega>0$, if $G_{k_\perp}(z_0,z_0)<0$.  In the long-wave limit
this is possible in the subsonic flow only and the corresponding
constraint coincides with Eq.~(\ref{attr}). More detailed
investigation shows that the potential, $G_\perp(z_0,z_0)$,  is
always positive if $\mu<1$. However, in the subsonic regime,
$\mu>1$,  there are regions of instability, that is,
$G_{k_\perp}(z_0,z_0)<0$ if  $0<q_\perp<q_{max}(a,\mu)$. The
latter are shown as shadowed areas in Fig.~\ref{fig3}. With the
increasing distance ($a\to\infty$) to the wall or decreasing
stream velocity ($\mu\to\infty$) the instability regions shrink to
zero, $q_{max}(a,\mu)\sim a^{-1/2},\;\mu^{-1/2}$.

\section{Conclusion}
To summarize, we have shown that the presence of the conducting
wall may drastically change the electrostatic interaction of the
dust grains in an ion flow. In particular, the electrostatic image
of the grain wake field may result in attraction between grains
levitating at the same height that, in its turn, yields the
Jeans-type instability of the dust layer.

It would be unduly naive  to draw quantitative conclusions from
the present calculations. However, it seems reasonable that even
in a real plasma sheath, which is essentially non-uniform, the
electrostatic image of the grain wake field may also affect the
motion of another grains outside the Mach cone. Although in most
experiments the screened Coulomb interaction is observed, there
are indications that the intergrain potential may be more
complicated. It was recently reported that, under certain
conditions, a void, \textit{i.e.,} a  dust-free region, appears in
a central part of a single dust layer \cite{tue}. The emergence of
a two-dimensional void in a layer consisting of some
 hundreds of grains can hardly be explained in the same manner as
a three-dimensional void; the latter requires strong influence of
the dust component upon the discharge structure \cite{goree}.
Although currently one cannot exclude that some additional
external forces appeared in the experiment \cite{tue}, we can
conjecture  that the 2D void formation is provided by complicated
intergrain interaction, for example, the one described in this
paper.

\section*{Acknowledgements}

 This study was supported in part
by the Russian Foundation for Basic Research (project no.
02-02-16439) and the Netherlands Organization for Scientific
Reseach (grant no. NWO 047.008.013).

\end{document}